\documentclass{aa} 

\usepackage{epsfig}
\usepackage{graphicx}
\usepackage{txfonts}
\usepackage[normalem]{ulem}
\usepackage[draft]{hyperref}

\graphicspath{{../../../Auxilliary/}}

\begin{document}

\title{Kernel-phase analysis: aperture modeling prescriptions that minimize calibration errors}

\author{Frantz Martinache\inst{1}, Alban Ceau\inst{1}, Romain Laugier\inst{1}, Jens Kammerer\inst{2,3}, Mamadou N'Diaye\inst{1}, David Mary\inst{1}, Nick Cvetojevic\inst{1} \& Coline Lopez\inst{1}}

\institute{Université Côte d'Azur, Observatoire de la Côte d'Azur, CNRS, Laboratoire Lagrange, France
  \and European Southern Observatory, Karl-Schwarzschild-Str 2, 85748, Garching, Germany \and Research School of Astronomy \& Astrophysics, Australian National University, ACT 2611, Australia}
 

\abstract
{Kernel-phase is a data analysis method based on a generalization of the notion of closure-phase invented in the context of interferometry, but that applies to well corrected diffraction dominated images produced by an arbitrary aperture. The linear model upon which it relies theoretically leads to the formation of observable quantities robust against residual aberrations.}
{In practice, detection limits reported thus far seem to be dominated by systematic errors induced by calibration biases not sufficiently filtered out by the kernel projection operator. This paper focuses on the impact the initial modeling of the aperture has on these errors and introduces a strategy to mitigate them, using a more accurate aperture transmission model.}
{The paper first uses idealized monochromatic simulations of a non trivial aperture to illustrate the impact modeling choices have on calibration errors. It then applies the outlined prescription to two distinct data-sets of images whose analysis has previously been published.}
{The use of a transmission model to describe the aperture results in a significant improvement over the previous type of analysis. The thus reprocessed data-sets generally lead to more accurate results, less affected by systematic errors.}
{As kernel-phase observing programs are becoming more ambitious, accuracy in the aperture description is becoming paramount to avoid situations where contrast detection limits are dominated by systematic errors. Prescriptions outlined in this paper will benefit any attempt at exploiting kernel-phase for high-contrast detection.}


\titlerunning{Kernel-phase: aperture modeling prescriptions}
\authorrunning{F. Martinache \& KERNEL team}

\maketitle


\section{Introduction}
\label{sec:intro}

Within the anisoplanetic field of an imaging instrument, in the absence of saturation, an in-focus image $I$ can formally be described as the result of a convolution product

\begin{equation}
I = O \star \mathrm{PSF}
\label{eq:convol}
\end{equation}

\noindent
between the spatially incoherent brightness distribution of an object $O$ and the instrumental point spread function (PSF). The careful optical design of telescopes and instruments assisted by adaptive optics (AO) attempts to bring the PSF as close as possible to the theoretical diffraction limit. Yet even for high quality AO correction, subtle temporal instabilities in the PSF make it difficult to solve for important problems such as: the identification of faint sources or structures in the direct neighborhood of a bright object (the high-contrast imaging scenario) or the discrimination of sources close enough from one another to be called non-resolved (the super-resolution scenario). Weak signals of astrophysical interest compete with time-varying residual diffraction features that render the deconvolution difficult.

The overall purpose of interferometric processing of diffraction-dominated images is to provide an alternative to the otherwise ill-posed image deconvolution problem. The technique takes advantage of the properties of the Fourier transform, that turns the convolution into a multiplication. One must however abandon the language describing images and instead, manipulate the modulus, also refered to as the visibility, and the phase of their Fourier transform counterpart. This Fourier transform can be sampled over a finite area of the Fourier plane traditionally described using the $(u,v)$ coordinates, whose extent depends on the geometry of the instrument pupil.

Non-redundant masking (NRM) interferometry uses a custom aperture mask featuring a finite number of holes that considerably simplifies the interpretation of images. Accurate knowledge of the mask's sub-aperture locations unambiguously associates particular complex visibility measurements in the image's Fourier transform to specific pairs of sub-apertures forming a baseline. The Fourier phase $\Phi$ at the coordinate $(u,v)$ is the argument of a single phasor:

\begin{eqnarray}
\phi(u,v) &=& \mathrm{Arg}\bigl(v_0(u,v) e^{i(\phi_0(u,v) + \Delta\varphi(u,v))}\bigr)\\
 &=& \Phi_0(u,v) + \Delta\varphi(u,v),
\end{eqnarray}

\noindent
where $v_0(u,v)$ and $\phi_0(u,v)$ respectively represent the intrinsic target visibility modulus and phase for this baseline, and $\Delta\varphi(u,v)$ the instrumental phase difference (aka the piston) experienced by the baseline at the time of acquisition. The same geometrical knowledge also makes it possible to combine together complex visibility measurements by baselines forming closing-triangles which lead to the formation of closure-phases: observable quantities engineered to be insensitive to diffferential piston errors affecting the different baselines.

Closure-phase was first introduced in the context of radio interferometry by \citet{1958MNRAS.118..276J} and eventually exploited in the optical starting with \citet{1986Natur.320..595B}. This useful observable enables NRM interferometry to detect companions at smaller angular separations than coronagraph can probe.

Kernel-phase analysis attempts to take advantage of the same property without requiring the introduction of a mask. The description of the full aperture requires a more sophisticated model that will reflect the intrinsically redundant nature of the aperture. Any continuous aperture can be modeled as a periodic grid of elementary sub-apertures resulting in a virtual interferometric array where every possible pair of sub-apertures forms a baseline. Whereas the NRM ensures that each baseline is only sampled once, the regular grid results in a highly redundant scenario. For a baseline of coordinate $(u,v)$ and redundancy $R$, the Fourier-phase will be that of the sum of $R$ phasors all measuring the same $\phi_0(u,v)$ but experiencing different realisations of instrumental phase $(\Delta\varphi_k)_{k=1}^R$:

\begin{equation}
\phi(u,v) = \mathrm{Arg}\bigl(\sum_{k=1}^R v_0(u,v) e^{i(\phi_0(u,v) + \Delta\varphi_k)}\bigr).
\label{eq:phasor}
\end{equation}

In the low-aberration regime provided by modern adaptive optics (AO) systems the impact residual pupil aberration $\varphi$ has on the Fourier phase can be linearized and Eq. \ref{eq:phasor} rewritten as:

\begin{equation}
\phi(u,v) = \phi_0(u,v) + \frac{1}{R} \sum_{k=1}^R \Delta\varphi_k.
\end{equation}

The list of what pairs of sub-apertures contribute to the complex visibility of a redundant baseline is kept in the {\it baseline mapping matrix} $\mathbf{A}$. It contains as many columns as there are sub-apertures ($n_A$) and as many rows as there are distinct baselines ($n_B$). Elements in a row of $\mathbf{A}$ are either $0$, $1$, or $-1$ (see Fig. 1 of \citet{2010ApJ...724..464M}). The phase sampled at all relevant coordinates of the Fourier-plane, gathered into a vector $\Phi$ can thus be written compactly as:



\begin{equation}
\Phi = \Phi_0 + \mathbf{R}^{-1} \cdot \mathbf{A} \cdot \varphi,
\end{equation}

\noindent
where $\mathbf{R}$ is the diagonal (redundancy) matrix that retains the tally of how many sub-aperture pairs contribute to the Fourier phase for that baseline. $\Phi_0$ is the Fourier phase associated with the object being observed (it is related to the object function $O$ of Eq. \ref{eq:convol} by the Van-Cittert Zernike theorem), and $\varphi$ is the aberration experienced by the aperture.
The redundancy $\mathbf{R}$ is expected to be directly proportional the modulus transfer function (MTF) of the instrument. The product $\mathbf{R}^{-1} \cdot \mathbf{A}$, referred to as the {\it phase transfer matrix}, describes the way pupil phase aberration propagate into the Fourier plane. The baseline mapping and the phase transfer matrices are rectangular and feature $n_B$ rows (the number of baselines) for $n_A$ columns (the number of sub-apertures in the pupil), with $n_B > n_A$.

As shown in \citet{2010ApJ...724..464M}, selected linear combinations of the rows of the phase transfer matrix will cancel the effect of the pupil phase $\varphi$. These linear combinations, gathered into a operator called $\mathbf{K}$ (the left-hand null space or kernel of the phase transfer matrix) project the Fourier-phase onto a sub-space that is theoretically untouched by residual aberrations. The resulting observables, called kernel-phases, are a generalisation of the concept of closure-phase, that can be found for an arbitrary pupil, regardless of how redundant.

Practice suggests that kernel- and closure-phase do not perfectly self-calibrate. Recently published studies using kernel-phase to interpret high-angular resolution diffraction dominated observations indeed lead to contrast detection limits mostly constrained by systematic errors \citep{2019MNRAS.486..639K, 2019A&A...623A.164L, 2019JATIS...5a8001S} instead of statistical errors \citep{2019A&A...630A.120C}. The goal of this paper is to provide insights into the limitations of Fourier-phase methods in general and to introduce the means to improve on these limitations.

This difficulty affects the kernel-phase interpretation of images as well as that of NRM interferograms. For despite the generalized assistance of adaptive optics during NRM observations \citep{2006SPIE.6272E.103T}, the need for long integration times and the use of sub-apertures that are not infinitely small means that interferograms are not simply affected by a simple and stable piston but by a time varying higher order amount of aberration \citep{2013MNRAS.433.1718I}. Closure-phase thus acquired on a point source therefore rarely reach zero and are biased.

Nevertheless, even when evolving over time, if the spatial and temporal properties of the perturbations experienced by the system remain stable across the observation of multiple objects, the overall resulting bias is also expected to remain stable. It is therefore possible to calibrate the closure-phases acquired on a target of interest with those acquired on a point-source. Thus calibrated closure-phases have been used as the prime observable for the detection of low to moderate contrast companions \citep{2008ApJ...679..762K}. There is however a limit to the stability of the observing conditions when hopping from target to target: changes in elevation, apparent magnitude for the adaptive optics, and telescope flexures, will result in an evolution of the closure-phase bias. Observations are therefore in practice never perfectly calibrating and the evolution of the calibration bias results in what is generally referred to as {\it systematic error}.

For low-to-moderate contrast detections up to a few tens, systematic errors are often a minor contribution that do not significantly affect the interpretation of the data. However as observing programs become more ambitious, attempting the direct detection of higher contrast companions \citep{2012ApJ...745....5K} in a part of the parameter space that cannot yet be probed by coronagraphic techniques, systematic errors become more important than statistical errors and one must resort to more advanced calibration strategies using multiple calibrators \citep{2013MNRAS.433.1718I, 2019MNRAS.486..639K}, to better estimate the calibration bias that will minimize the amount of systematic error.

\section{Fourier-phase calibration errors}
\label{sec:calib}

\begin{figure}
\includegraphics[width=\columnwidth]{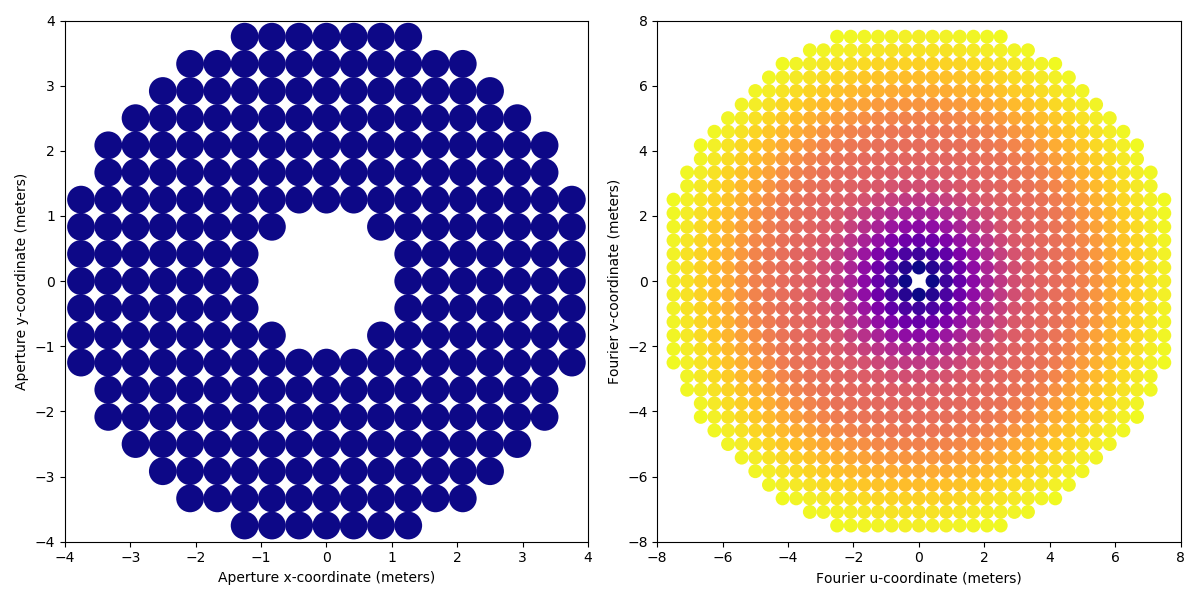}
\caption{Binary discrete representation of the SCExAO instrument pupil for kernel-phase analysis. Left panel: the dicretized instrument pupil built from a square grid of pitch $s$ = 42 cm. Right panel: the Fourier coverage associated to this discretization. The color code used to draw the Fourier coverage reflects the redundancy associated to the virtual interferometric baselines.}
\label{f:subaru_42cm_bina_model}
\end{figure}

Kernel-phase analysis requires to approximate the near-continuous aperture of the telescope by a discrete representation: a virtual array of sub-apertures, laid on a regular grid of predefined pitch $s$, paves the surface covered by the original aperture. Computation of all the possible pairs of virtual sub-apertures in the array leads to the creation of a second grid of virtual baselines, the majority of which are highly redundant. An example is shown in Fig.  \ref{f:subaru_42cm_bina_model} for the aperture of an 8-meter telescope, discretized with a grid of pitch $s = 42$ cm. Keeping track of what pairs of sub-apertures contribute to all the baselines leads to the computation of the baseline mapping matrix $\mathbf{A}$ and the redundancy matrix $\mathbf{R}$. The two are used to eventually determine the kernel operator $\mathbf{K}$ that generalizes the notion of closure-phase, as introduced by \citet{2010ApJ...724..464M}.

\begin{figure}
\includegraphics[width=\columnwidth]{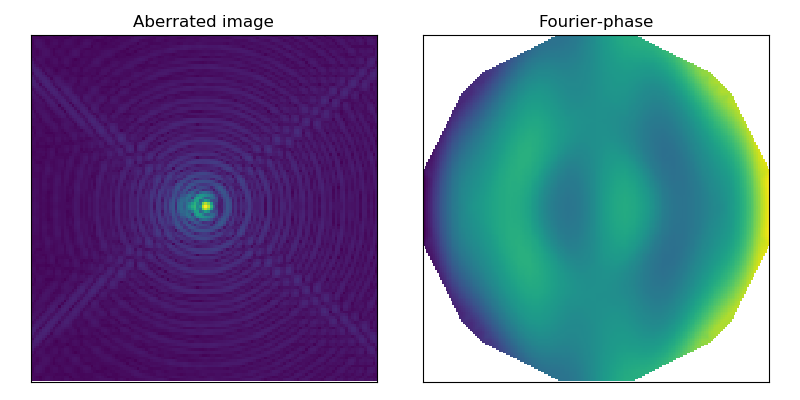}
\caption{Left: Simulated monochromatic ($\lambda$ = 1.6 $\mu$m) SCExAO image of a 10:1 binary in the presence of 100 nm of coma along the axis of the binary. Right: the associated Fourier-phase ranging from $\pm$1.5 radian (see also Fig. \ref{f:effect}).}
\label{f:ex_img}
\end{figure}

\begin{figure}
\includegraphics[width=\columnwidth]{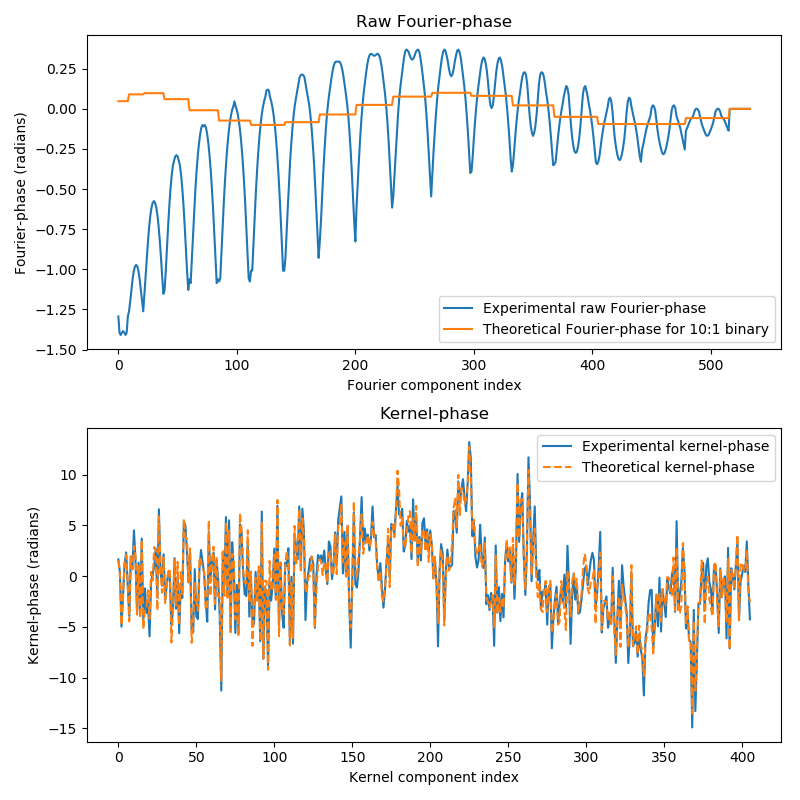}
\caption{Demonstration of the impact of the kernel processing. The top panel shows that the experimental Fourier phase extracted from a single aberrated image shown in Fig. \ref{f:ex_img} (the blue curve) bears little ressemblance with the theoretical expected binary signal (in orange). Contrasting with the raw Fourier-phase, the bottom panel shows how the projection onto the kernel susbpace efficiently erases the impact of the aberration and brings the experimental kernel-phase curve ($\mathbf{K} \cdot \Phi$), also plotted using a solid blue line, much closer to its theoretical counterpart ($\mathbf{K} \cdot \Phi_0$), now plotted using a dashed orange line so as to better distinguish them.}
\label{f:effect}
\end{figure}

The following simulation will illustrate the interest of this line of reasoning. Using a single, simulated, monochromatic ($\lambda = 1.6\,\mu$m) and noise-free image of a 10:1 contrast binary object (located two resolution elements to the left of the primary) affected by a fairly large (100 nm rms) amount of coma, shown in the left panel of Fig. \ref{f:ex_img}. The Fourier-phase $\Phi$ extracted from this image (shown in the right panel of Fig. \ref{f:ex_img}) appears to be completely dominated by the aberration. The plot of the same raw Fourier-phase (the blue curve in the top panel of Fig. \ref{f:effect}) compared to the predicted Fourier signature of the sole binary $\Phi_0$ confirms this observation. Using the kernel operator $\mathbf{K}$ computed according to the properties of the discrete model\footnote{The model is computed using a python package called XARA, developped in the context of the KERNEL project, available for download \url{http://github.com/fmartinache/xara/}} represented in Fig. \ref{f:subaru_42cm_bina_model}, it is possible to project the 546-component noisy Fourier phase vector $\Phi$ onto a sub-space that results in the formation of a 414-component kernel. The bottom panel shows how despite the drastic difference between the raw and theoretical Fourier-phase, the two resulting kernels match one another: the kernel operator effectively erases the great majority of the aberrations affecting the phase present in the Fourier space while leaving enough information to describe the target in a meaningful manner, such that:

\begin{equation}
  \mathbf{K} \cdot \Phi = \mathbf{K} \cdot \Phi_0.
\end{equation}

Although quite satisfactory in its apparent ability to reduce the impact of the aberration, the match of the kernel curves is not perfect. The small difference between the two example curves is what is generally refered to as the calibration error. This error can be independantly measured using a second image, this time of a single point source (a calibrator), assuming that the system suffers the same aberration. In this noise-free scenario, subtraction of the kernel-phase extracted of one such calibration image would result in a perfect match. An instrumental drift between the two exposures would result in a new bias. If the magnitude of this new bias becomes comparable to or larger than the statistical uncertainties, the interpretation of kernel- and closure-phase typically requires to invoke a tunable amount of systematic error added in quadrature to the uncertainty.

\section{Kernel-phase discretization prescriptions}
\label{sec:strat}

Given that no noise was included in this ideal scenario, the fact that some aberration leaks into the kernel and results in the need for a calibration suggests that the discrete model used to describe how pupil phase propagates into the Fourier plane is not sufficiently accurate and we will look into ways to improve it.

\subsection{Building a discrete representation}
\label{sec:build}

\begin{figure}
  \includegraphics[width=0.7\columnwidth]{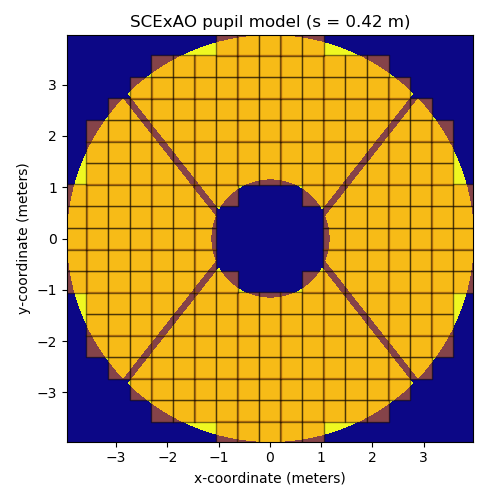}
  \caption{Example of discretized version of SCExAO's pupil using a square grid, aligned with the center of the aperture. Only the virtual sub-apertures for which the transmission, determined as the normalized intersection between the virtual sub-aperture and the underlying pupil, is greater or equal to 50 \% are kept as valid sub-apertures contributing to the model. The $s = 0.42$ m pitch value is chosen so as to fit an entire number (here 19) of sub-apertures across the pupil diameter.}
    \label{f:grid_match}
\end{figure}

The discretization process is as follows: a high-resolution 2D image of the aperture is generated from the details of the pupil specifications (outer and inner diameter, spider thickness, angle and offset). A square grid of sub-apertures of pitch $s$ is laid atop the pupil image and points falling within the open parts of the aperture are kept in the model. To be counted amongst the virtual sub-apertures, the area of the transmissive part of the original aperture overlapping with the region covered by the square virtual sub-aperture has to be greater than 50 \%.
When building the model, it will be critical to ensure that no virtual baseline is greater than the actual telescope diameter: this would indeed result in attempting to extract information outside the Fourier domain the true aperture gives access to. To mitigate this risk, one will first ensure that an entire number of sub-apertures fits within one aperture diameter, and then eliminate all the computed baselines greater than the aperture diameter.

A regular grid is required so that the density of the discrete representation is as homothetic as possible to the original aperture: this translates into a model redundancy $\mathbf{R}$ that better matches the true modulation transfer function (MTF) of the instrument. It is also important to align the grid with the aperture model so that the symmetry properties of the apertures are reflected in its discrete representation: one either uses a grid that is centered on the aperture and that features an integer odd number of apertures (the option retained to build the discretized aperture shown in this paper) or an offset grid with an integer even number of sub-apertures.

Fig. \ref{f:grid_match} introduces the example what will serve as the reference to compare the relative merits of different discrete models. It uses the Subaru Telescope pupil mask of the SCExAO instrument \citep{2015PASP..127..890J}, characterized by its large (2.3 m diameter) central obstruction and non-intersecting thick spider vanes at the non-trivial angle of 51.75 degrees \citep{2009PASP..121.1232L}. This non-trivial aperture geometry makes it a rich test case. Using the aforementionned recommendations, a centered grid with a $s = 0.42$ m pitch, fits almost exactly 19 virtual sub-apertures across the aperture nominal diameter of 7.92 meters.

The $n_A = 272$ virtual sub-apertures of this array form $n_B = 562$ distinct baselines. As discussed in \citet{2013PASP..125..422M}, for a rotational symmetric of order 2 aperture\footnote{The aperture is identical to itself rotated by 180$^\circ$ relative to its center}, the Fourier-phase and its kernels are insensitive to even order aberrations. This property reflects in the properties of the linear phase transfer model: the number of non-singular values of the baseline transfer matrix $\mathbf{A}$ should be equal to $n_E = n_A / 2$, therefore leaving $n_K = n_B - n_A / 2$ kernel-phases. For the kernel analysis to lead to optimal results, it is important to ensure that these properties are verified. An aperture that does not respect this symmetry condition will in contrast result in less ($n_K = n_B - n_A + 1$) kernels.

\subsection{Grid pitch and image size}
\label{sec:pitch}

The 42 cm pitch of the grid illustrated in Fig. \ref{f:grid_match} does not offer enough resolution to reflect the presence of the 25 cm thick spider vanes of the real aperture. This contrasts with models that have generally been used since the inception of kernel-phase (see for instance Fig. 2 of \citet{2010ApJ...724..464M}) that have naively overemphasized the impact of spider, which in turn contributes as will be made clear below, to amplify the calibration bias.

The pitch $s$ of the grid is of course a free parameter that can be adjusted: the finer the grid, the more representative of the details of the pupil it is expected to become and the more capable of capturing higher spatial frequency components of the images. A discrete model with a finer pitch however implies the description of a wider effective field of view (of radius $0.5 \lambda/s$) over which the kernel-analysis can lead to meaningful results. The size of the image will therefore impose a limit to how fine the pitch can get.

For the wavelength ($\lambda = 1.6\,\mu$m) of the simulations used in this Section, the plate-scale (16.7 mas per pixel) and size ($128 \times 128 $) of the images suggest that the pitch cannot be finer than $s > 206.265 \times 1.6 / (128 \times 16.7) \approx 0.15$ m. Beyond this simulation scenario, image noises induced by dark current, readout and photon noise, and a preference for computationally manageable problems will guide the user toward using coarser models in practice.

\begin{figure*}
  \includegraphics[width=\textwidth]{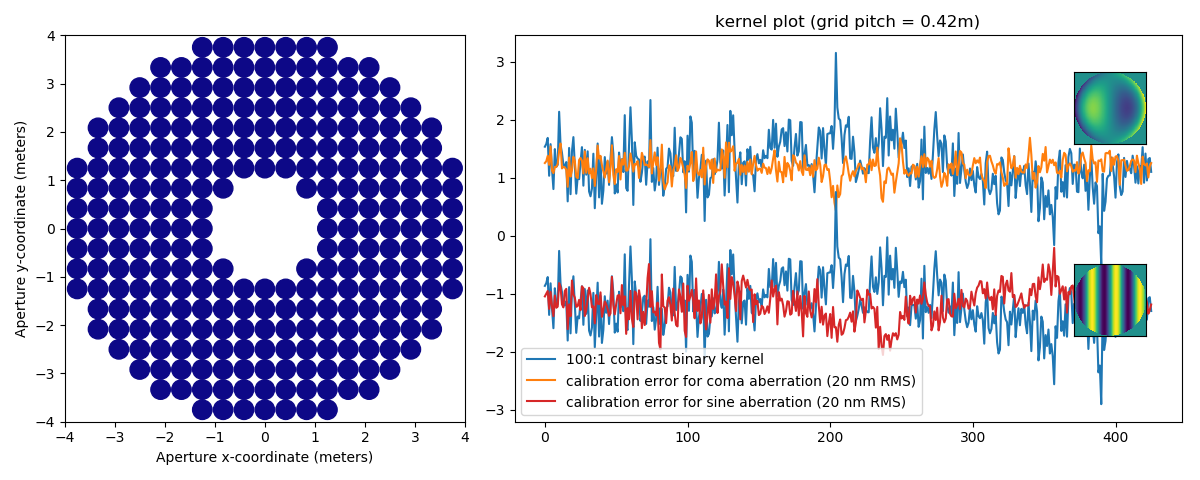}
  \includegraphics[width=\textwidth]{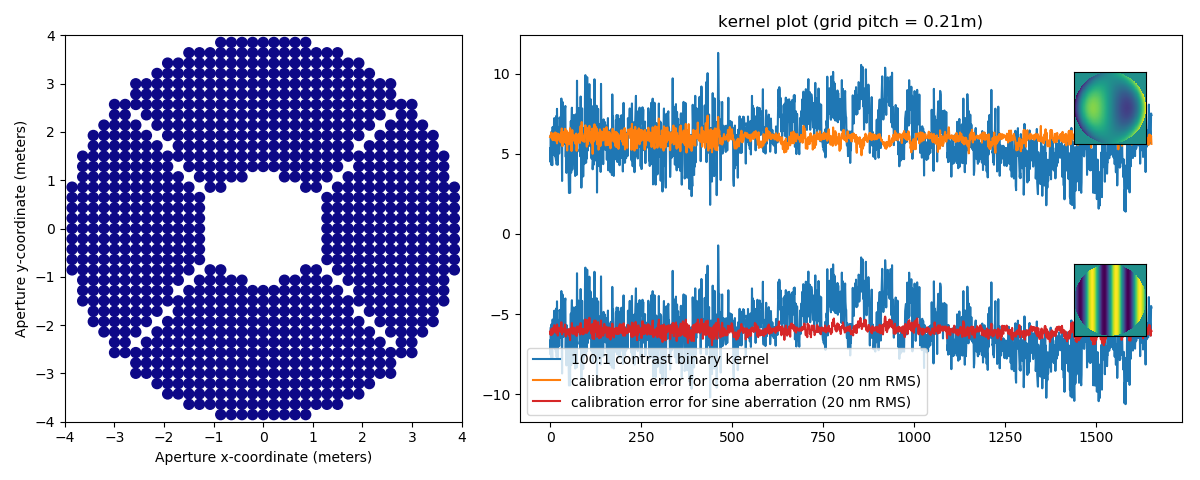}
  \includegraphics[width=\textwidth]{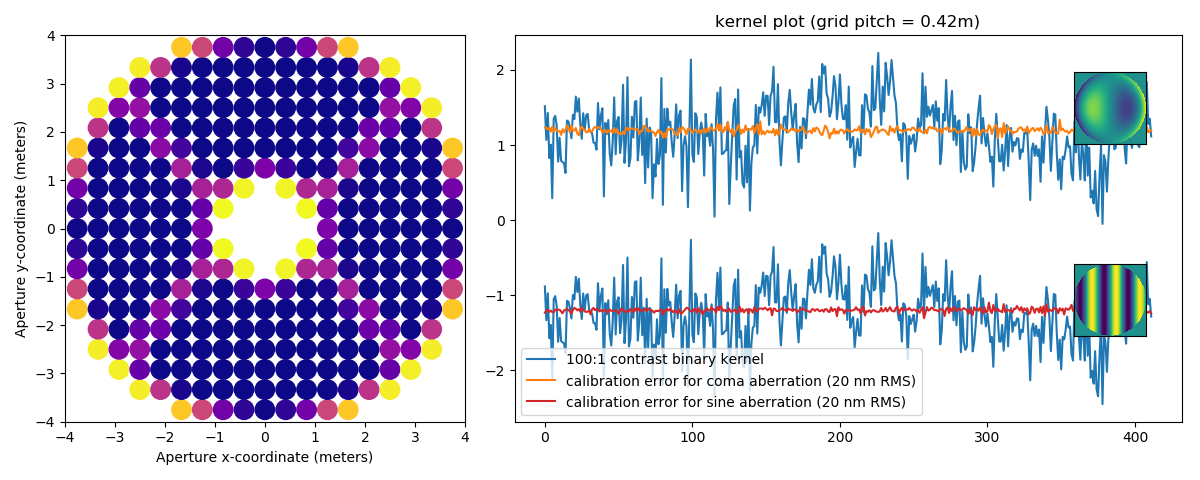}
\caption{Comparison of the self-calibrating performance of the kernel-phase analysis of a single image for three discrete models of the same aperture. Each of the three panels features, side by side, a 2D representation of the discrete aperture model used and a plot of the kernels extracted from the image of a point source (the calibration error), in the presence of either coma (the orange curve) or a 3-cycle sinusoidal aberration (the red curve) and how they compare to the signal of a 100:1 contrast binary (the blue curve). The top panels presents the reference binary model of the SCExAO pupil, with a 42 cm pitch; the middle panel, a denser model with a 21 cm pitch that more accurately matches the fine structures of the telescope; the third panel, a model using the original 42 cm pitch grid but including the transmission function.}
\label{f:comp_models}
\end{figure*}

\subsection{Comparing models}
\label{sec:comp}

To assess the relative merits of multiple models, we look at the impact the discretization strategy has on the magnitude of the calibration bias. We have seen that the pitch of the model impacts the overall dimension of the problem. It also impacts the associated redundancy $\mathbf{R}$ and therefore the overall magnitude of the kernel-phases extracted from a given image. To enable a meaningful comparison of multiple models, we will compare the rms of the calibration bias to the theoretical standard deviation of the theoretical signature (see for instance eq. 27 of \citet{2019A&A...630A.120C}) induced by a 100:1 contrast companion that would be located two resolution elements to the left of the primary along the horizontal axis.

The simulations systematically include a 20 nm rms static (odd) aberration: either a 3-cycle horizontal sinusoid or coma along the same direction. These two examples were selected because they are both perfectly odd, and therefore have full impact on the analysis, and because they feature different structures: the impact of the sinusoidal modulation is more uniformly distributed across the aperture, whereas the impact of the coma (like that of most higher order Zernike modes) is stronger toward the edges.
The same two images ($128 \times 128$ pixels, one featuring coma and one featuring the sinusoidal aberration) are processed using the kernel-phase pipeline, using three discrete models. The results of these three analysis are summarized in Fig. \ref{f:comp_models} featuring, side by side, a rendering of the discrete aperture model and the plot of thus biased kernel-phase vector extracted from either image, and in Table \ref{tbl:perf} summarizing the dimensions of the models and their intrinsic sensivitity to calibration error.

\begin{table}
\begin{tabular}{l | c | c | c }
  & sparse binary & dense binary & sparse grey \\
  \hline
  pitch  & 0.42 & 0.21 & 0.42 \\
  $n_A$ & 272 & 956 & 300 \\
  $n_B$ & 562 & 2132 & 562 \\
  $n_K$ & 426 & 1654 & 412 \\
  \hline
  ref. signal & 0.435 & 1.583 & 0.402 \\
  coma bias & 0.180 (41 \%) & 0.317 (20 \%) & 0.036 (9 \%) \\
  sine bias & 0.286 (66 \%) & 0.263 (17 \%) & 0.026 (6 \%) \\
  \hline
\end{tabular}
\caption{Summary of the model properties and their performance. $n_A$, $n_B$ and $n_K$ respectively represent the number of sub-apertures, the number of baselines and the number of kernels of each model. The coma and sine bias rows respectively show the magnitude of the bias induced 20 nm rms of coma and the sinusoidal aberration, in radians}
\label{tbl:perf}
\end{table}

The first model, presented in the top panel of Fig. \ref{f:comp_models} is the reference using the $s=0.42$ m pitch grid introduced earlier. Using this model, the magnitude of the calibration bias extracted from the images affected by either type of aberration represents a significant fraction (of the order of 50 \%) of the signature of the 100:1 binary companion. Under such circumstances, the contrast detection limits associated to these uncalibrated kernel-phase are likely to be rapidly dominated by this systematic error. The middle panel of Fig. \ref{f:comp_models} illustrates the impact of a finer $s=0.21$ m grid pitch: the model better reflects the presence of the spiders and the overall shape of the pupil. A larger number of kernels is extracted from the same image (almost four times as many) but more importantly for this discussion, the relative magnitude of the calibration bias is reduced by a factor $\approx 2-3$: a kernel-phase analysis based on a finer and more accurate description of the original aperture will feature reduce model-induced calibration errors and will therefore be less susceptible to calibration errors in general.

Increasing the resolution of the grid is not the only option available. One can indeed also refine its description by allowing for a variable sub-aperture transmission. In addition to deciding whether to keep or discard one virtual sub-aperture as part of the model, the information on the clear fraction of the sub-aperture, translated into a local transmission value can be appended and taken into account when creating the phase transfer model. Such a ``grey aperture'' model makes it possible to more accurately describe the edges and high-spatial frequency features of the aperture without necessarily increasing the pitch of the model. One example using such a continuous transmission model is illustrated in the bottom panel of Fig. \ref{f:comp_models}: despite using a grid pitch identical to the reference model, the discrete representation of the aperture clearly better outlines the finer features of the aperture as the trace of the spider vanes becomes visible. For this example, the transmission cut-off value was set to 10$^{-3}$: the grey model includes a slightly higher number of virtual sub-apertures than its binary counterpart for which the cut-off was set to 0.5. In the end, one forms a number of kernels (see the Appendix for a general discussion regarding the number of kernels) similar to the binary case. The improvement brought by the inclusion of this transmission model is substantial: the magnitude of the bias is brought well below 10 \% that of the signature of the 100:1 binary companion.

The two effects of a finer resolution and a transmission model can be compounded to lead to even better performance. Generally, whether one uses a binary or a grey model, doubling the resolution of the grid leads to an improvement by a factor $\approx$2. The performance of kernel-phase reaches a point where the details of the implementation of the upstream simulation becomes critical.

Overall, there seems to be no significant difference between the two types of aberrations introduced. Sinusoidal modulation seem to be better processed in general, likely because of the sharper edge structure of the coma, that systematically requires more resolution. The impact of aberrations of higher spatial frequency, beyond what the chosen model can effectively describe are filtered out either by adequate image cropping (following the recommendations given in Sec. \ref{sec:pitch}) or by application of an image mask. We can conclude that including the aperture transmission model is a major improvement that renders the kernel-phase analysis less susceptible to systematic errors.

\section{Kernel-phase analysis revisit}
\label{sec:applications}

In this section, we use the recommendations outlined in the previous section and apply them to a series of datasets whose kernel-phase analysis has already been published. We will feature two applications: the analysis of a ground-based dataset published by \citet{2016MNRAS.455.1647P} and an extended version of the dataset used for the original kernel-phase publication by \citet{2010ApJ...724..464M}. The review of these two applications will further illustrate the importance of better aperture modeling practices for kernel-phase analysis.

\subsection{Palomar/PHARO}
\label{sec:p3k}

The data consists of two data-cubes of 100 images of the binary system $\alpha$-Ophiuchi \citep{2011ApJ...726..104H} and of the single star $\epsilon$-Herculis that were acquired with the PHARO instrument \citep{2001PASP..113..105H} at the focus of the Palomar Hale Telescope, after AO correction provided by the P3K AO system \citep{2013ApJ...776..130D}. 

The data-cubes were recovered from an archive linked in the original publication. The preprocessed large $512 \times 512$ pixels original frames were first cropped down to a more manageable $64 \times 64$ pixel size. With a plate scale of 25 mas per pixel, the field of view is still $\pm$800 mas. To reach sufficient resolution in the Fourier space, a fast Fourier transform (FFT) based extraction algorithm such as the one used in the original study requires an adapted amount of zero-padding. The now standard complex visibility extraction method of XARA instead explicitly computes the discrete Fourier transform for the spatial frequencies of the discrete model, such as suggested by \citet{2007OExpr..1515935S} and filters out sub-pixel centering errors as used by \citet{2019MNRAS.486..639K}: the cropping of the image not only filters out the higher level of noise brought out by weakly illuminated pixels, it also is more computationally efficient as it requires the computation of smaller Fourier transform matrices.

\begin{figure}
\includegraphics[width=\columnwidth]{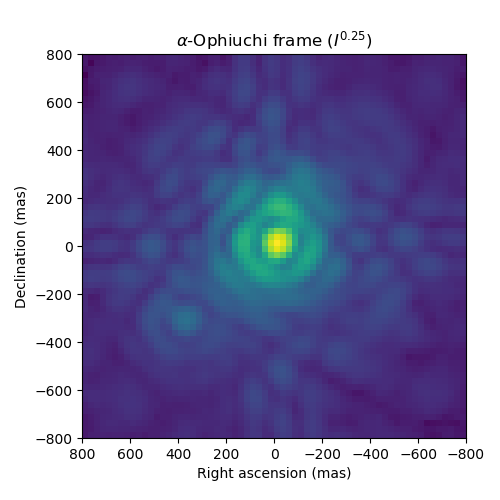}
\caption{Example of frame acquired on $\alpha$-Ophiucus (non-linear scale: power 0.25). Notable features of this image include the thick and slightly tilted diffraction spikes induced by the medium-cross pupil mask used at the time of acquisition, and the ghost induced by the neutral density filter in the bottom left quadrant of the frame. The companion later recovered by the kernel-phase data reduction is buried underneath the first diffraction ring, to the left of the primary.}
\label{f:p3k_frame}
\end{figure}

Images were acquired using the K$_S$ filter (central wavelength: 2.14 $\mu$m) and the medium cross pupil mask inside the PHARO camera was used to limit the risk of saturation in the image for the otherwise bright target of interest. An example of image is shown in Fig. \ref{f:p3k_frame}. The images presents a few noteworthy features: the apparent companion clearly visible in the bottom left quadrant is a ghost induced the 0.1 \% neutral density filter used at the time of the acquisition. This ghost is present in all the frames, including those acquired on the calibrator. Also visible are strong diffraction spikes induced by the very thick spider vanes of the medium-cross mask, whose orientation does not quite match the grid of pixels (upper vertical spike leans slightly to the left).

\begin{figure}
  \includegraphics[width=\columnwidth]{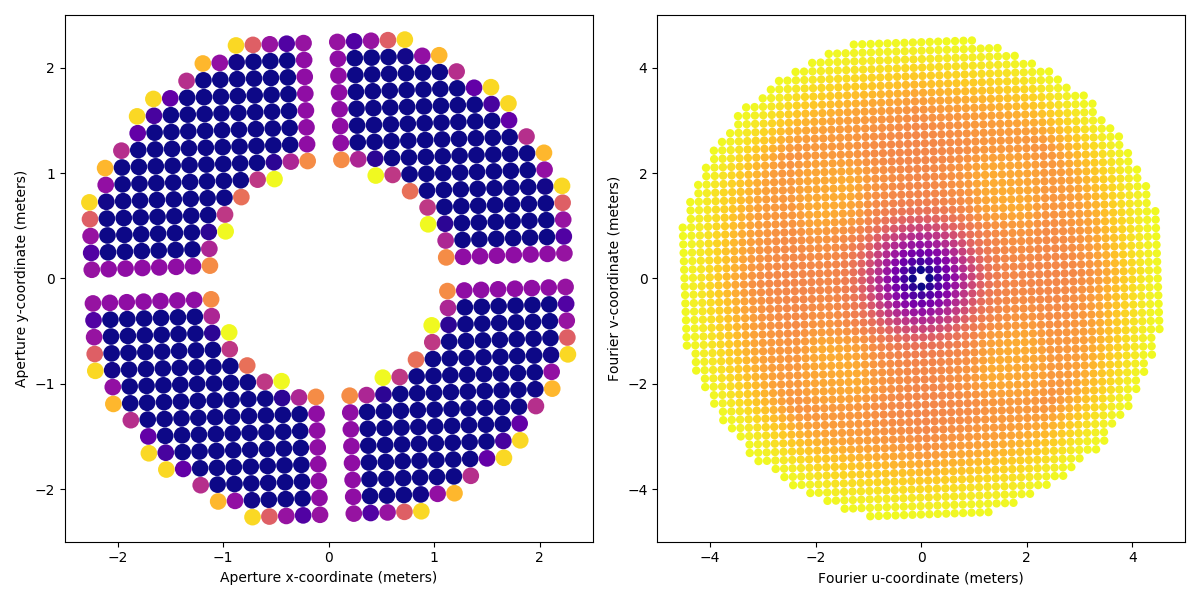}
  \caption{Representation of the discrete model (left: aperture, right: Fourier coverage) of the PHARO med-cross aperture, using the transmission model of the true aperture. To further reduce the amount of systematic error, the model was built using a square grid that was rotated to match the orientation of the original pupil mask. The impact of the presence of the spiders in the model is revealed in the Fourier plane as four small depressions appear in the intermediate spatial frequency range.}
  \label{f:pharo_model}
\end{figure}

We built a new discrete grey model of the medium-cross based on the specifications published by \citet{2001PASP..113..105H} that were confirmed by an image of the pupil enabled by one of the modes of the camera. In the image provided by the pupil imaging mode of the PHARO camera, the medium-cross mask appears to be rotated counterclockwise by two degrees. We used the recipe outlined in Sec. \ref{sec:strat} to produce grey discrete representation of the true aperture  using a square grid with a pitch $s = 0.16$ meters that was then rotated to match the grid to the true aperture. To eliminate possible mistakes, we used a simulation reproducing the properties of the PHARO K$_S$-band images that included a fixed amount of aberration and rotated our mask until we found the orientation that minimizes the amount of calibration error. The optimum model thus identified is shown in Fig. \ref{f:pharo_model}.

The presence of the ghost in all the images will contribute to the calibration bias of the data. \citet{2016MNRAS.455.1647P} chose to further window the data so as to mask the ghost out before attempting to extract the kernels: this however leaves too few useful pixels to lead to the formation of $n_K = 1048$ kernels of the model ($n_A = 528$, $n_B = 1312$). In this analysis, we keep all the information available in the image, under the assumption that the contribution of the ghost will be erased when subtracting the kerrnel-phase from the calibrator.

In the high-contrast regime (which in practice applies when the contrast is greater than 10:1), the amplitude of the kernel-signature of a binary is expected to be directly proportional to the contrast (secondary/primary). This makes it convenient to compute for a grid of positions around the primary, the normalized dot product between the calibrated signal and the theoretical signal of a high-contrast binary computed for all grid positions. The use of such colinearity maps was introduced by \citet{2019A&A...623A.164L}: the presence of a clear maximum in this map shows where the input signal best matches the theoretical signature of a binary.

\begin{figure*}
  \includegraphics[width=0.33\textwidth]{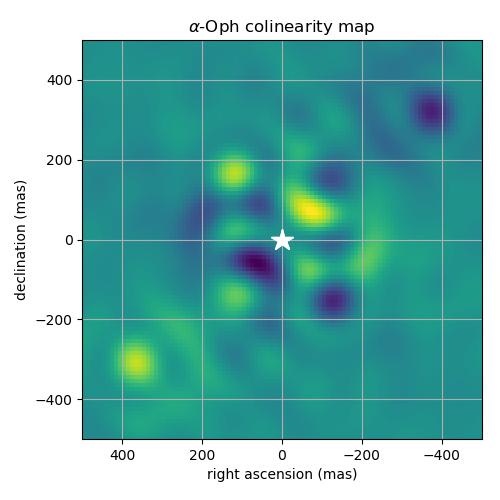}
  \includegraphics[width=0.33\textwidth]{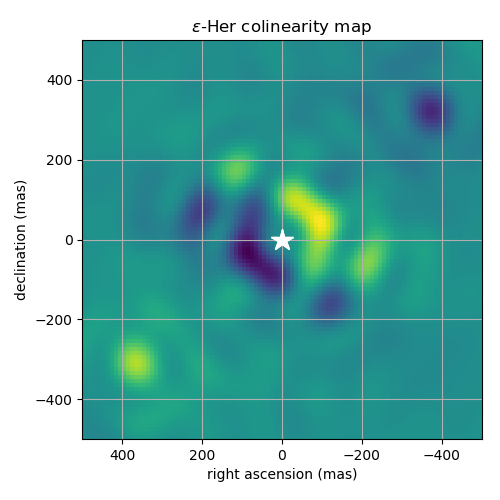}
  \includegraphics[width=0.33\textwidth]{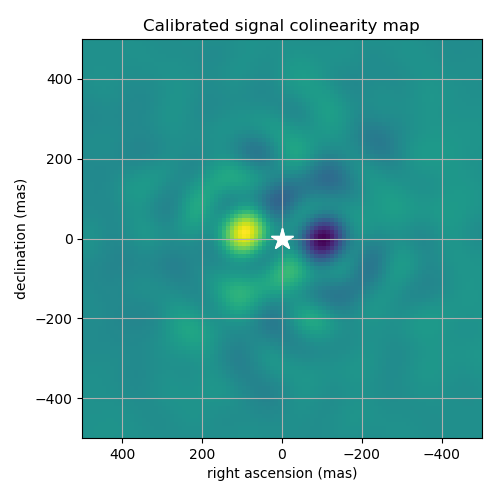}
  \caption{Colinearity maps for the raw and calibrated kernel-phases extracted from the data, over a $\pm$500 mas field of view. The left panel shows the map built from the uncalibrated kernel-phases of $\alpha$-Oph. The middle panel shows the same map built from the uncalibrated kernel-phase of $\epsilon$-Her. The right panel shows the colinearity map of the calibrated signal, that is the difference between the kernel-phases of $\alpha$-Oph and $\epsilon$-Her. The two uncalibrated map prominently feature the signature of the ghost present in all images in the bottom left quadrant as well as other structures at closer separations. The map of the calibrated signal on the other hand shows that most of these features are gone and reveals a positive bump on the left hand side of the central star (for a separation of 100 mas and a position angle of 84 degrees), taken as indication of the presence of a companion.}
\label{f:P3K_colin_maps}
\end{figure*}

Fig. \ref{f:P3K_colin_maps} shows the result of this computation for the raw signal of both target and calibrator as well as for the calibrated signal of $\alpha$-Ophiuchi over a $\pm$500 mas field of view. Kernel-phase is, like the closure-phase, a measure of asymmetry of a target so the colinearity map is always antisymmetric. The two uncalibrated map prominently feature the signature of the ghost present in all images in the bottom left quadrant as well as other structures at closer separations (up to $\sim$200 mas). Whereas the signature of the ghost is expected, these intermediate separation features (particularly on the map of the calibrator) suggest that the kernel-phases are still affected by a calibration bias. Our efforts have ensured that the modeling induced errors are minimal. however, given that individual images were integrated over 1.4 s, that is many times the coherence time, the kernel-phase are still affected by an additional bias induced by temporal variance described by \citet{2013MNRAS.433.1718I}. We can observe that the subtraction of the kernel-phases of the calibrator from those obtained on $\alpha$-Ophiucus effectively erases these features along with that of the ghost. The bright bump (and its antisymmetric dark counterpart) clearly visible to the left (and the right) of the central star in the right panel of Fig. \ref{f:P3K_colin_maps} is indicative of the quality of the detection of the companion around $\alpha$-Ophiuchi.

\begin{figure}
  \includegraphics[width=\columnwidth]{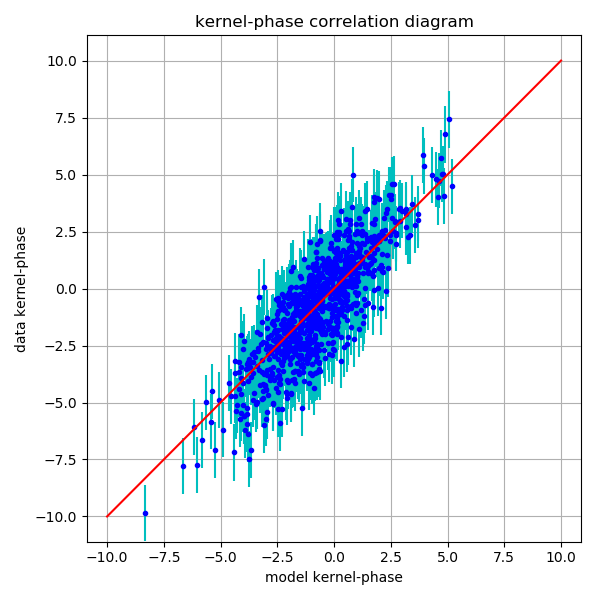}
  \caption{Correlation plot of the calibrated signal of $\alpha$-Ophiuchi with the best binary fit solution: following the hint provided by the colinearity map of the calibrated data, the signal present in the data is fairly well reproduced by a binary companion located at angular separation $\rho = 123.4 \pm 1.7$ mas, position angle $\theta = 86.5 \pm 0.5$ and contrast $c = 25.1 \pm 0.1$. Residual structure in the data is accounted for by the introduction of an ad hoc amount of systematic error so that the reduced $\chi^2 = 1$.}
  \label{f:P3K_fit}
\end{figure}

We use the location of the maximum of colinearity as the starting point for a traditional $\chi^2$ minimization algorithm using the Levenberg-Marquardt method that is shipped as part of the python package scipy. The uncertainties associated to the calibrated kernel-phases are simply computed as the quadratic sum $\sigma_e$ of the uncertainties deduced from variance between frames for the $\alpha$-Ophiuchi and $\epsilon$-Herculis. The result of this optimization is represented in the correlation plot of Fig. \ref{f:P3K_fit}: the model fit looks very convincing and locates the companion in the area hinted at in the calibrated colinearity maps of Fig. \ref{f:P3K_colin_maps}.

The careful modeling of the aperture unfortunately does not suffice to eliminate the need for ad hoc systematic errors at the time of the optimization: a relatively large amount of systematic error ($\sigma_s = 1.2$ rad) still needs to be added quadratically to the experimental dispersion ($\sigma_E = 0.1$) in order to give a reduced $\chi^2 = 1$, for the following parameters: separation $\rho = 123.5 \pm 2.9$ mas, position angle $\theta = 86.5 \pm 0.2$ degrees and contrast $c = 25.1 \pm 1.1$ . It should be pointed out that these don't quite match the values reported (see Table 1 of \citet{2016MNRAS.455.1647P}) for the NRM observation that usually set the standard. It should however also be pointed out that the new contrast estimate is in good agreement with measurements reported in Table 3 of \citet{2011ApJ...726..104H}, taken when the companion was at larger angular separation.

While the signature of the companion is more clearly visible in this analysis than in the results reported by \citet{2016MNRAS.455.1647P}, the situation is still not fully satisfactory, as our improved model of the aperture did not lead to a detection with uncertainties on the binary parameters driven solely by statistical errors. Several explanations were invoked in the original publication to justify the sub-par performance: they still apply here. The sub-standard seeing conditions that induce variability in the AO correction on targets of distinct magnitudes and the fact that both sources were acquired in very different areas of the detector explain in great part how the statistical variance experienced during the observation cannot on its own be representative of all errors affecting the kernel-phase. This new analysis, becauses it uses a model that is adapted to the information availabla in the data-cubes, nevertheless draws a more favorable picture for kernel-phase, which shows here a much more convincing result.

\subsection{HST/NICMOS}
\label{sec:hst}

As pointed out in Sec. \ref{sec:pitch}, the seminal kernel-phase publication used a rather crude discrete representation of the aperture of the Hubble Space Telescope and was nevertheless able to report the detection of a companion to the M-dwarf GJ 164 \citep{2009ApJ...695.1183M}. In attempting to accurately model the effective aperture of the NICMOS1 instrument used to acquire the data, we refered to the work of \citet{1997hstc.work..192K}. that suggests the presence of an important ($\sim$10 \%) misalignment of the instrument cold mask relative to the original optical telescope assembly (OTA) that was completely overlooked by \citet{2010ApJ...724..464M}.

Multiple datasets recovered from the HST archival were acquired on GJ 164 on epoch 2004-02-14 UT (proposal ID \#9749) in several narrow band filters: F108N, F164N and F190N \citep{2009nici.book.....V}. Our updated analysis also includes images of calibration star SAO 179809 observed at a single epoch (1998-05-01, proposal ID \#7232) acquired in the F190N filter. The original $256 \times 256$ pixel images were bad-pixel corrected, recentered and cropped down to $84 \times 84$ pixel size, over which the F190N filters seems to feature good SNR, before being gathered into data-cubes. With a plate scale of 43 mas per pixel, the effective field of view is thus $\pm$ 1.8 arc second. According to the image sampling constraints reminded in Sec. \ref{sec:pitch}, the size of the field of view and the wavelength of acquisition set a limit to how fine the pitch can get, which for the F190N filter translates into $s = 0.109$ m. Although the model pitch could be updated for the shorter wavelengths (which for a fixed image size give access to an increasing number of spatial frequencies), we build a single discrete model (including transmission) for a homogeneous analysis across the entire data-set that will enable comparison across spectral bandpasses.

\begin{figure}
  \includegraphics[width=\columnwidth]{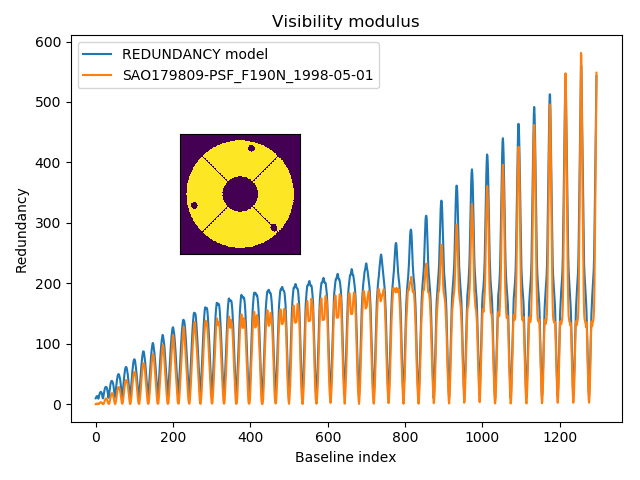}
  \includegraphics[width=\columnwidth]{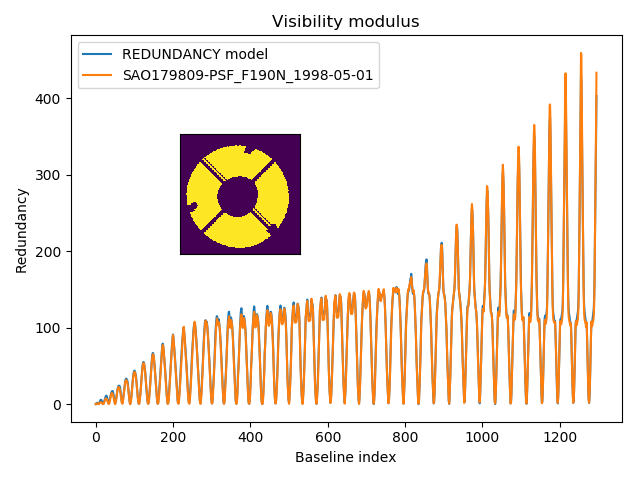}
  \caption{Comparison of the predicted redundancy with the experimental OTF (estimated from the modulus of the Fourier transform of F190N images of calibration star SAO 179809) for two models: the first (top panel) assumes that the aperture is that of the HST Optical Telescope Assembly (OTA) and the second (bottom panel) takes into account the misaligned cold mask of the NICMOS camera. Whereas the redundancy of the first model fails to reproduce the modulus of the Fourier transform effectively measured from the image, the second model convincingly matches the fine features of the instrument OTF.}
  \label{f:OTF}
\end{figure}

We however first need to demonstrate that the discrete model indeed benefits from the updated aperture description recommended by \citet{1997hstc.work..192K}. For this we use the images of the calibration star SAO 179809. Sec. \ref{sec:build} introduced the idea that an accurate discrete model should translate into a predicted redundancy diagonal matrix $\mathbf{R}$ that matches the true instrument OTF, which for a calibration star should correspond to the  modulus of the complex visibility extracted for the different baselines of the model. Fig. \ref{f:OTF} illustrates this property and compares the redundancy associated to models characterized by the same $s = 0.109$ m pitch for two apertures: one that includes the outline of the OTA only (top panel) and one that includes the misaligned NICMOS cold mask (bottom panel). Whereas the OTA model should already be an improvement over the one originally used, we can observe that the associated redundancy fails to reproduce the modulus of the Fourier transform computed for the corresponding spatial frequencies. The gap is particularly visible for intermediate spatial frequencies that feature less power than what is predicted by the model. The more accurate model including the misaligned cold mask is a major improvement as the predicted redundancy $\mathbf{R}$ almost perfectly matches the fine features (in particular the dropped lobes visible for baseline indices ranging from 400 to 800) of the experimental OTF.

\begin{figure}
\includegraphics[width=\columnwidth]{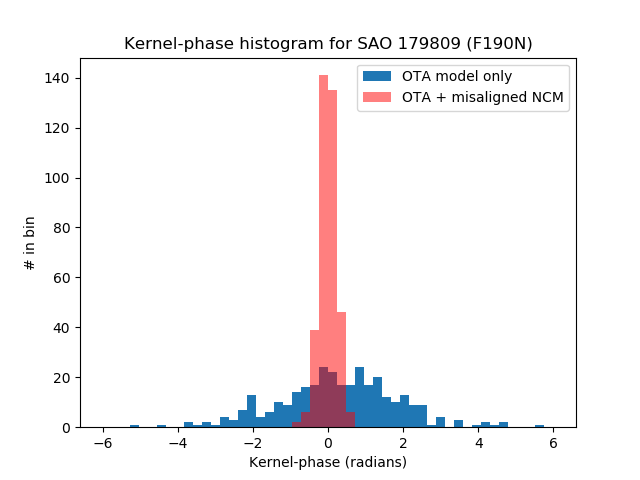}
\caption{Kernel-phase histograms computed from a set of images of calibration source SAO 179809 using two aperture models characterized by the same pitch $s = 0.109$ m. The first model assumes that the aperture is that of entire optical telecope assembly (OTA) and results in the blue histogram. The second model takes into account the misaligned cold mask inside the NICMOS camera and results in the red histogram.}
\label{f:KP_histo}
\end{figure}

Unlike any of the previously considered scenarios, the pupil is here clearly not rotation symmetric so we don't expect to form the optimal number of kernels (see the end of Sec \ref{sec:build}). The more accurate model is nevertheless expected to translate into more accurate kernel-phase. SAO 179809 being a calibration source, we can verify that the magnitude of the calibration biases decreases by comparing (see Fig. \ref{f:KP_histo}) the histograms of the kernel-phase computed using the two aforementioned models. The improvement is significant with a reduction of the standard deviation by a factor $\sim$10, despite a larger number of kernels in the better model (375 vs 320), demonstrating one more time that a more accurate model reduces the impact of calibration errors. With the accurate model, the magnitude of the calibration bias ($\sigma_S = 0.222$ radians) is now comparable to that of a 100:1 contrast ratio companion (rms = 0.215 radians) located two resolution elements away from the primary would give for this discrete model.

The images of SAO 179809 were acquired more than 5 years before those of GJ 164: although we will keep using the same aperture model, so we can't expect to use the kernel-phase of SAO 179809 reliably to calibrate the kernel-phase signal of GJ 164. The magnitude of the calibration error estimated from the observation of SAO 179809 can however provide an order of magnitude for the expected fitting error for a binary such as GJ 164.

\begin{figure}
  \includegraphics[width=\columnwidth]{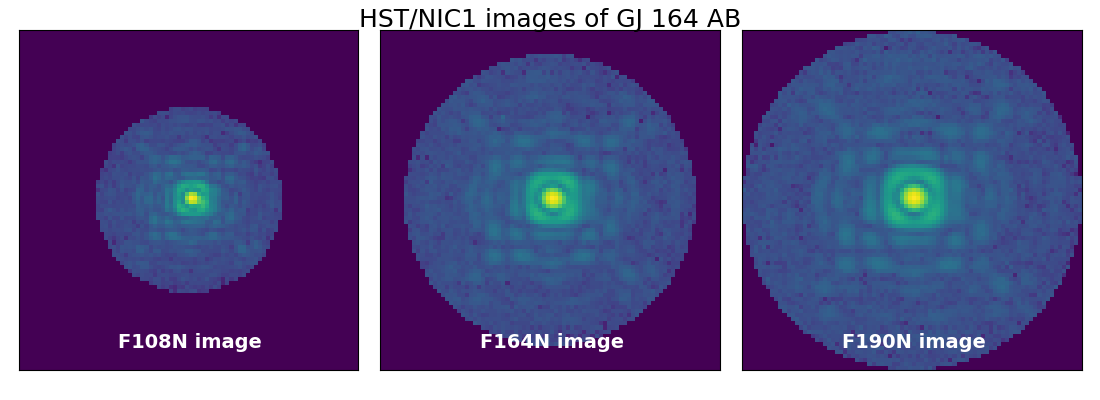}
  \caption{Snapshots of GJ 164 AB for the 2004-12-23 epoch in the three NICMOS narrow band filters: F108N, F164N and F190N. The non-linear scaling of the image makes the window applied to the data more apparent. Given that a single discrete model with a fixed pitch is used to process this data-set, the window, used to cover a finite number of resolution elements, must be scaled linearly with the wavelength. Here, the effect of the window is mostly to reduce the contribution of poorly illuminated pixels to the overall noise of the kernel-phase.}
  \label{f:gj164_imgs}
\end{figure}

One interesting feature of the GJ 164 dataset is the availability of images acquired in multiple filters: 80 in the F190N filter, 40 in the F164N filter and 10 in the F108 filter for this 2004-12-23 epoch. Fig. \ref{f:gj164_imgs} shows a snapshot of GJ164 for these three filters. In addition to the expected linear scaling of the diffraction with the wavelength, one will observe the linear scaling size of the circular window matching the effective field of view induced by the choice of a unique model with a fixed pitch. The phase transfer model at the core of the kernel-analysis is achromatic: pupil coordinate points and baselines are indeed all expressed in meters and not in radians as is customary in long baseline interferometry. At the time of data extraction however, the wavelength needs to be taken into account in the computation of a discrete Fourier transform matrix to match the sampling of the data.

The published analysis of the F190N images has revealed that a companion is present at an angular separation $\sim$90 mas, which is of the order of $0.5 \lambda/D$. In the high-contrast regime, the kernel-phase signature of a binary companion of contrast $c$ (primary / secondary) at wavelength $\lambda$ has a simple analytic expression:

\begin{equation}
  \mathbf{K} \cdot \Phi_0(u,v) = \mathbf{K} \cdot \frac{1}{c} \times \sin{\frac{2\pi}{\lambda} (\alpha u + \beta v)},
\end{equation}

\noindent
where $\alpha$ and $\beta$ are the angular Cartesian coordinates of the companion (in radians), $u$ and $v$ are vectors collecting the coordinates of the baselines (in meters). Assuming that the contrast of the companion is constant for the different filters, we can explicit the derivative of the binary kernel-phase signal relative to the wavelength:

\begin{equation}
  \frac{\partial}{\partial\lambda}\mathbf{K} \cdot \Phi_0(u,v) = \mathbf{K} \cdot \frac{1}{c} \times \cos{\frac{2\pi}{\lambda} (\alpha u + \beta v)} \times \frac{-2\pi}{\lambda^2}.
  \label{eq:scaling}
\end{equation}

\begin{figure}
  \includegraphics[width=\columnwidth]{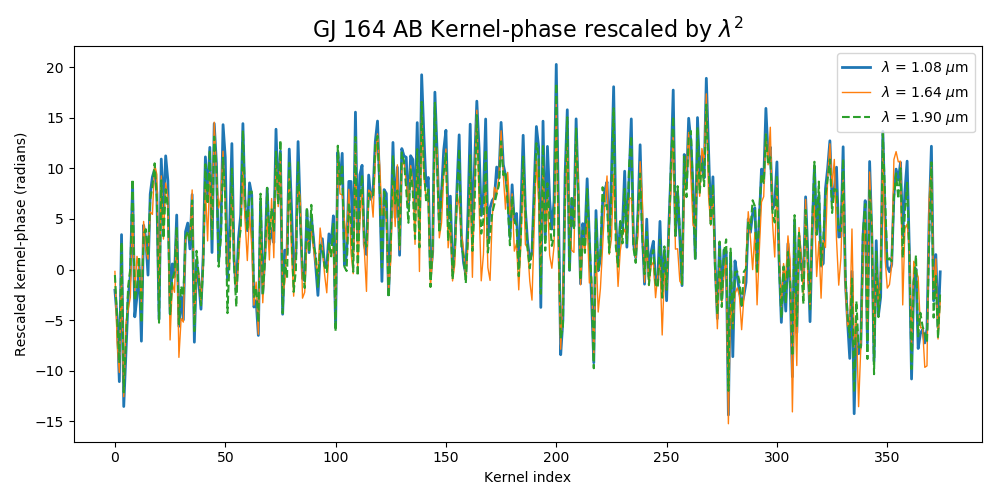}
  \caption{Representation of the median kernel-phase vector extracted for the three filters data-sets (F108N, F164N and F190N), rescaled by the wavelength of the filter (taken in microns) squared. Thus rescaled, the three signals are very consistent with one another, confirming the presence of a near constant contrast structure partly resolved from the central star.}
  \label{f:gj164_scaled_kernels}
\end{figure}

If the companion is unresolved, the cosine term varies slowly and the dominant wavelength dependant effect is the overall $1/\lambda^2$ scaling factor of Eq. \ref{eq:scaling}. Thus by multiplying kernel-phase extracted in the different filters by associated wavelength (expressed in microns) squared, we expect to see signals of comparable structure and magnitude. Fig. \ref{f:gj164_scaled_kernels} shows the result of one such comparison for the median signal extracted from the three sets of images. The stability of the signature of the companion across the three bands (covering almost a decade) is striking, suggests that the contrast is indeed stable for the different filters, and attests of the consistency of the kernel-phase data analysis.

\begin{figure*}
\includegraphics[width=\textwidth]{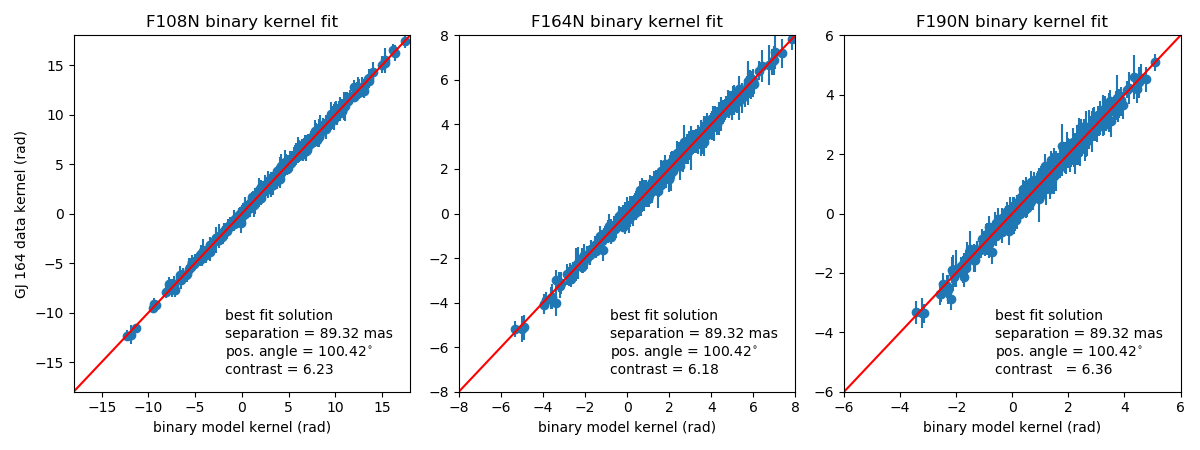}
\caption{Correlation plots for the combined 5-parameter model-fit of the multi-filter GJ 164 AB image data-set, split back between the different filters (from left to right: F108N, F164N and F190N). The detection of the companion by the kernel-phase analysis is unequivocal.}
\label{f:gj164_correl_plots}
\end{figure*}

Going from 1.9 to 1.08 $\mu$m however almost doubles the resolving power: the signature of the companion, expected to be degenerate in the F190N filter, for which one observes a strong correlation between contrast and angular separation, will be better constrained by the F108N observation. The three data-sets are combined to feed a five parameter model fit optimization algorithm: two astrometric parameters and three contrasts, for a total of 1120 degrees of freedom. The result of this global optimization is represented in the correlation plots of Fig. \ref{f:gj164_correl_plots}, split by filter. The best solution places the companion 89.3 $\pm$ 0.4 mas away from the primary at the 100.4 $\pm$ 0.1 degree position angle , and leads to the following three contrasts: 6.23 $\pm$ 0.1 in the F108N filter, 6.19 $\pm$ 0.1 in the F164N filter and 6.36 $\pm 0.1$ in the F190N filter. Fig. \ref{f:gj164_correl_plots} also shows that the $1/\lambda^2$ signal scaling factor of the binary signal (see Eq. \ref{eq:scaling}) leads to intrinsically higher signal to noise ratio for the observation at the shorter wavelength.

Although the astrometric solution for the combined fit is generally consistent with the result published by \citet{2010ApJ...724..464M}, the contrast in the F190N is revised and drops from $9.1 \pm 1.2$  to $6.36 \pm 0.1$. The origin of the initial overestimation of the contrast, not captured by the uncertainty, is not clear and can likely be attributed to a combination of causes: the use of an innapropriate aperture model in the first place, the strong contrast-separation degeneracy of the F190N observation and an overall more mature data analysis process today. One can incidentally note that the revised F190N and F164N contrasts are in better agreement with the majority of the contrasts measurements  reported by \citet{2009ApJ...695.1183M} with NRM observations using broad band H and K filters. 

In the absence of a calibrator, the computation of parameter uncertainties required the introduction of a controlled amount of systematic error (added in quadrature to the measured statistical uncertainties) so that the reduced $\chi^2$ is unity. In this case, $\sigma_S = 0.15$ rad amount of systematic error is required, which seems to comparable to the magnitude of the calibration bias that was estimated ($\approx$0.22 rad) after the analysis of the SAO 179809 data-set. Unlike what was the case with the PHARO dataset (see Sec. \ref{sec:p3k}) it seems our modeling of the aperture and the interpretation of the resulting data meets our expectations.

\section{Discussion}
\label{sec:disc}

While we were able to show that the modeling prescriptions outlined in this paper do bring closure- and kernel-phase closer to the true self-calibration, it seems that in order to reach the highest contrast detection limits one will always resort to calibration observations, which typically require telescope repointing and is therefore a time consuming option. If a target were to exhibit different properties at two nearby wavelengths, such as a strong absorption or emission spectral line caught by one filter and not the other, it seems a powerful and more efficient calibration scheme would be to subtract the kernel-phase acquired in the two filters from one another. One would then have to fit the thus calibrated data to a spectral differential kernel-phase model.

\begin{figure} \includegraphics[width=\columnwidth]{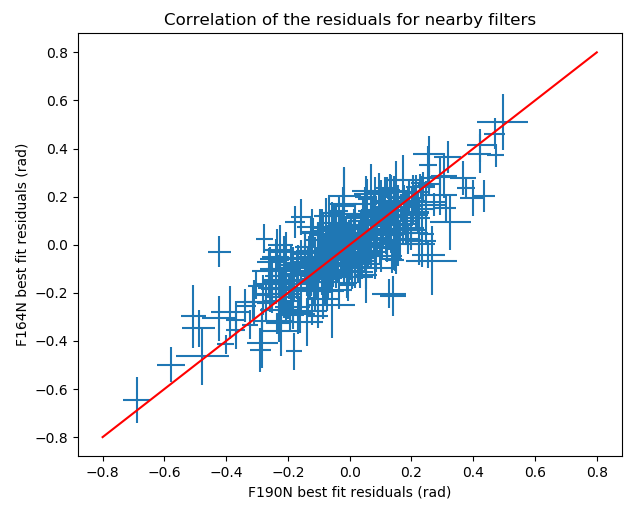}
  \caption{Correlation plot of the kernel-phase residuals after subtraction from the best-fitting binary model. The relatively good match between the two residuals (correlation coefficient $\approx$ 0.87) suggests that the use of spectral differential kernel-phase would be a valid way to solve the calibration problem.}
  \label{f:residuals}
\end{figure}

We have seen that with its stable contrast, GJ 164 AB does not really feature any noteworthy spectral behavior and that filters are reasonably far apart from one another so this dataset is not ideal to try this idea out. Nevertheless, because of the relative proximity of the F164N and F190N filters, we can still assess the potential of one such observing mode by looking for similarities in the structure of the kernel-phase residuals after subtraction of the best fit model. Fig. \ref{f:residuals} features a correlation plot of these residuals for the F164N and F190N filters that include experimental uncertainties. The apparent good correlation between the two residuals suggests that a spectral-differential calibration scheme has some merit: the magnitude of the differential kernel-phase residual is $\sim0.11$ rad, which is getting close to the associated experimental dispersion ($\sigma_E = 0.08$ rad). This approach should be further tested on images acquired in two filters less further away, or in the analysis of data-cubes produced by an AO-fed integral field spectrograph, which will be the object of future work.


\section{Conclusion}
\label{sec:conclu}

Kernel-phase is a versatile adaptation of the idea of closure-phase that can be used in a wide variety of configurations, assuming that images are reasonably well corrected. With versatility however comes the need for care. The description of the aperture upon which the analysis is made must be thought through, requiring good knowledge of the original pupil and matched to the constraints brought by images, in particular the number of useful pixels, as well as the scientific ambition.

We have seen that the inclusion of a transmission model for the description of the aperture required to build the pupil-Fourier phase transfer model brings a major improvement in fidelity. Several examples using ideal numerical simulations and actual data-sets from ground-based observations as well as from space have demonstrated that this overall higher fidelity reduces the impact of systematic errors and leads to better results. One should also note that the introduction of grey transmission model now makes it possible to take advantage of pupil apodization masks used to reduce the contribution of photon noise over a finite area of the image, which, assuming good self-calibration, will result in improved contrast detection limits.

Closure- and kernel-phase based observing programs are becoming more and more ambitious with instruments that make it theoretically possible in some cases to probe for planetary mass companions \citep{2019JATIS...5a8001S, 2019A&A...630A.120C} down to the diffraction limit without a coronagraph. The proper handling of systematic errors in both scenarios is becoming paramount. While efficient calibration procedures offer a way to recover from problematic solutions, the work described here is an attempt to avoid them in the first place. 

\begin{acknowledgements}
This project has received funding from the European Research Council (ERC) under the European Union's Horizon 2020 research and innovation program (grant agreement CoG \# 683029). It is based in part on observations made with the NASA/ESA Hubble Space Telescope, obtained from the data archive at the Space Telescope Science Institute. STScI is operated by the Association of Universities for Research in Astronomy, Inc. under NASA contract NAS 5-26555.
\end{acknowledgements}

\appendix

\section{Number of kernels for a symmetric aperture}

Given the properties of the expected theoretical number of kernels $n_K$ associated to the model of a circular obstructed aperture, one can compute an approximate number of expected kernels, based only on its dimension and the pitch $s$ of the grid used to build the model. Neglecting the effect of spider vanes, the total number of virtual sub-apertures $n_A$ that fit within a circular aperture of diameter $D$ featuring a central obstruction of diameter $d$ is proportional to the surface area of the aperture: $S_A = \pi \times (D^2 - d^2) / 4$. The number of baselines $n_B$ will itself be proportional to the surface area of the Fourier coverage, which is half of a disk of diameter $2D$: $S_{UV} = \pi D^2$.

If the aperture is 180-degree rotational symmetric, we know \citep{2013PASP..125..422M} that the number of kernels $n_K$ yielded by the model will be $n_K = n_B - n_A/2$. We can therefore compute the ratio of the number of kernels and the  number of baselines to find a number that only depends on the geometric properties of the aperture $r = n_K/n_A = 1 - 0.25 \times (1 - (d/D)^2)$. This makes it possible to evaluate the kernel-efficiency of a circular aperture: of the order of 77 \% for the SCExAO pupil case discussed in the body of the paper.

Given the step $s$, one can estimate how many virtual baselines will fit within the Fourier coverage: $n_B = S_{UV} / s^2$ and predict the number of kernels one can build, using the ratio determined above: $n_K = r \times n_B$. Applications for the reference model ($s = 0.42$ m) predict $n_{K} = 430$ kernels, and for the finer model ($s = 0.21$ m), $n_K = 1719$, which are reasonably close to the true values listed in Table \ref{tbl:perf}.

\bibliographystyle{aa}

\end{document}